\documentclass[11pt]{article}
\usepackage{fullpage}
\usepackage{lipsum} 

\usepackage[sc]{mathpazo} 
\usepackage[T1]{fontenc} 
\usepackage{microtype} 

\usepackage[hmarginratio=1:1,top=32mm,columnsep=20pt]{geometry} 
\usepackage{multicol} 
\usepackage[hang, small,labelfont=bf,up,textfont=it,up]{caption} 
\usepackage{booktabs} 
\usepackage{float} 
\usepackage{hyperref} 
\usepackage{amsmath}
\usepackage{paralist} 
\usepackage{epsfig} 

\usepackage{abstract} 

\usepackage{titlesec} 
\renewcommand\thesection{\Roman{section}} 
\renewcommand\thesubsection{\Roman{subsection}} 
\titleformat{\section}[block]{\large\scshape\centering}{\thesection.}{1em}{} 
\titleformat{\subsection}[block]{\large}{\thesubsection.}{1em}{} 

\usepackage{fancyhdr} 

\title{\vspace{-15mm}\fontsize{24pt}{10pt}\selectfont\textbf{Adaptive Control of 4-DoF Robot manipulator}} 

\author{
\large
\textsc{Pavel Mironchyk}\thanks{University Of Washington}\\[2mm] 
\normalsize \href{mailto:p.mironchyk@yahoo.com}{p.mironchyk@yahoo.com} 
\vspace{-5mm}
}
\date{}

\begin{document}

\maketitle 

\thispagestyle{fancy} 


\begin{abstract}

\noindent In experimental robotics, researchers may face uncertainties in parameters of a robot manipulator that they are working with. This uncertainty may be caused by deviations in the manufacturing process of a manipulator, or changes applied to manipulator in the lab for sake of experiments. Another situation when dynamical and inertial parameters of a robot are uncertain arises, is the grasping of objects by a manipulator. In all these situations there is a need for adaptive control strategies that would identify changes in dynamical properties of manipulator and adjust for them. This article presents a work on designing of an adaptive control strategy for 4-DoF manipulator with uncertain dynamical properties, and outcomes of testing of this strategy applied to control of simulator of robot.
\end{abstract}


\begin{multicols}{2} 

\section{Introduction}

This article presents an adaptive control strategy for 4-DoF manipulator, where some dynamical parameters of a manipulator are considered not to be known precisely. Also it is considered that there is no way to measure these parameters other than start controlling and trying to determining precise values of parameter while manipulator executing some tasks. So, the proposed controlling strategy must have integrated online recursive parameter identification. 

To derive the strategy, first the kinematics of manipulator is explained. The kinematics is necessary to derive then inverse dynamics equations of manipulator, which are then used then to implement adaptive torque controller. The final section of this article is dedicated to simulation and testing of the proposed strategy.

There are number of adaptive control strategies described in literature (see [Lewis, 2011]). But the distinguishing feature of the proposed strategy is that it focuses on identification of parameters that are related to position of centers of masses in manipulator bodies.  

\section{Strategy}

\subsection{Kinematics}

A normal way to describe robot kinematics is to use Danavit and Hartenberg parameters. The given robot manipulator represent an anthropomorphic leg. Four degrees of freedom are distributed in the following way three revolute joints combined into one in foundation of the \"hip\", and one revolute joint in the \"knee\". The three combined revolute joints act as spherical joint in hip of a human for instances. In real robotic manipulators it would be possible to \"merge\" all three joints into points, and they have to be separated by some physical links. But since this the target of the work was to simulate adaptive control strategy,and merging all of them into one simplifies derivation of robot dynamics greatly, it was decided to merge them into 3-dof z-y-x joint with zero length links. 
Table 1 contains the DH parameters of the robot. Further the straightforward equations of robot kinematics are given.

\end{multicols}

\begin{table}[H]
\centering
\footnotesize\setlength{\tabcolsep}{20.5pt}
\begin{tabular}{l@{\hspace{24pt}} *{4}{c}}
\toprule
\bfseries $Link_{i}$ & \multicolumn{4}{c}{\bfseries Parameters} \\
& $\theta$ & $d$ & $a$ & $\alpha$ \\
\midrule
\bfseries 1
& 0 & 0 & ~0 & $\frac{\pi}{2}$ \\
\bfseries 2
& 0 & 0 & ~0 & $\frac{\pi}{2}$ \\
\bfseries 3
& 0 & 0 & $L_{upper leg}$ & $\frac{\pi}{2}$ \\
\bfseries 4
& 0 & 0 & $L_{lower leg}$ & $\frac{\pi}{2}$ \\
\bottomrule
\addlinespace
\end{tabular}
\caption{DH-parameters of robot kinematics}\label{tab:dhparams}
\end{table}


\[
  r_{upper body}=R_z(q_{z_{hip}})R_y(-q_{y_{hip}})R_x(q_{y_{hip}})r_{CoM upper body}
\]
\[
  r_{lower body}=R_z(q_{z_{hip}})R_y(-q_{y_{hip}})R_x(q_{y_{hip}})[\begin{bmatrix} 
0  \\
0  \\
L_{upper body} 
\end{bmatrix} + R_y(q_{y_{knee}})r_{CoM_{lower body}}]  
\]
\[
  r_{tip}=R_z(q_{z_{hip}})R_y(-q_{y_{hip}})R_x(q_{y_{hip}})[\begin{bmatrix} 
0  \\
0  \\
L_{upper body} 
\end{bmatrix} + R_y(q_{y_{knee}})\begin{bmatrix} 
0  \\
0  \\
L_{lower body} 
\end{bmatrix}]  
\]

\begin{multicols}{2} 
Where $R_x$,$R_y$,$R_z$ are rotational matrices. 

To illustrate kinematics of manipulator, a visualization of manipulator in some sample configuration is given below.
\epsfig{file=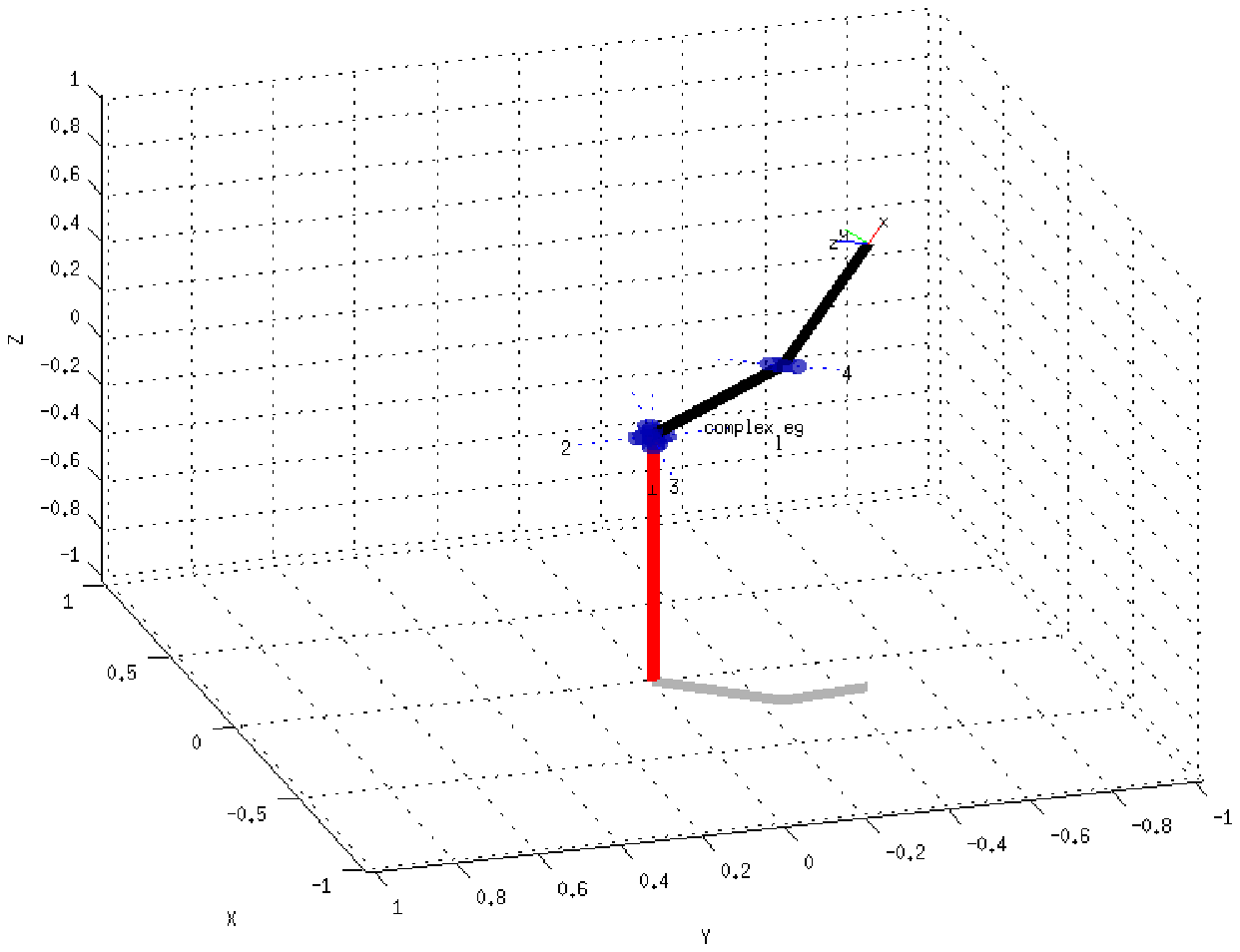, width=3.0in}

\subsection{Dynamics}

Robot dynamics is derived using Lagrangian-Euler mechanics. Lagrange's equation of motion for a conservative system are given by
\[
\frac{d}{dt}\frac{\partial L}{\partial \dot{q}} - \frac{\partial L}{\partial q} = \tau
\]
where $L = K - P$, $\tau$ is a vector of generalized forces, $q$ is a joint variables vectors, $K$ is kinetic energy of the system, $P$ is a potential energy.
A more standard form of manipulator dynamics is 
\[
M(q)\ddot{q} + C(q,\dot{q}) + G(q)=\tau
\]
where $M(q)$ is the inertia matrix, $C(q,\dot{q})$ is Coriolis/centripetal vector, and G(q) is the gravity vector.

For the given kinematics of a robot the Lagrangian can be derived by 
\[
  L=K_{upper body} + K_{lower body} - P_{upper body} - P_{lower body}
\]
Where
\[
   K = \frac{1}{2}mass{\frac{\partial r}{\partial q}}^{T}*\frac{\partial r}{\partial q}
\]
And 
\[
   P = mass {\begin{bmatrix} 
0  \\
0  \\
g 
\end{bmatrix}}^{T} r
\]

At the end 
\[
\frac{d}{dt}\frac{\partial L}{\partial \dot{q}} - \frac{\partial L}{\partial q} = W(q,\dot{q},\ddot{q}) = \tau
\]

\subsection{Parametrization}
To implement parameter identification algorithm, the manipulator dynamics has to be parametrized linearly. The goal of this project to design a strategy that would adapt to uncertain model parameters, which are center of masses of linked bodies in manipulator. It is not possible to factor out positions if center of masses from the inverse dynamic of a manipulator directly, as the dynamics equation is non-linear in relation to coordinates of center of masses. However it is possible multiple these nonlinear terms to some parameters and through this was to adjust inverse dynamics equation of the system. 
So then $W$ can be factored as
\[
W(q,\dot{q},\ddot{q})=\Phi(q,\dot{q},\ddot{q})\theta^{'}
\]
Where 
\[
\theta^{'}=\begin{bmatrix} {L_{CoM_{upperleg}}}^2 \\
L_{CoM_{upperleg}} \\
{L_{CoM_{loweleg}}}^2 \\
L_{CoM_{upperleg}} \\
1
\end{bmatrix}
\]
Lets introduce set of parameters 
\[
\theta = \begin{bmatrix} 
a \\
b \\
c \\
d \\
e
\end{bmatrix}
\]
So we can parametrize $W(q,\dot{q},\ddot{q})$ as 
\[
W^{'}(q,\dot{q},\ddot{q})=\Phi(q,\dot{q},\ddot{q})(\theta^{'}\cdot\theta)= {\Phi^{'}(q,\dot{q},\ddot{q})}\theta
\]
So the parametrarized inverse dynamics looks like:
\[
{\Phi^{'}(q,\dot{q},\ddot{q})}\theta=\tau
\]

\subsection{Computed Torque Controller}

In foundation of the described control strategy lays a computed torque controller. A idea behind a computed torque controller is to cascade non-linear system with its inverse so the overall system has constant unity gain. In practice the inverse is not perfect so a feedback loop is required to deal with errors.
The computed torque controller is given by:
\[
{\Phi^{'}(q + K_p  e ,\dot{q} + K_d \dot{e},\ddot{q})}\theta=\tau
\]
Where $K_{p}$ and $K_{v}$ are position and velocity gains matrices, $e=q^{*}-q$ is position error.

\subsection{Parameters Identification}

The recursive least squares algorithm is used for parameter estimation. As it is the easiest way most reliable way for parameter estimation in the proposed control strategy. 
\[
W^{'}(q,\dot{q},\ddot{q})={\Phi^{'}(q,\dot{q},\ddot{q})}\theta  
\]
Or at some $i$-th step:
\[
W^{'}_i={\Phi^{'}_{i}}{\theta}_{i}={\tau}_{i}
\]
So the RLS scheme can be expressed in the following steps:
\begin{equation}
\hat{\theta_{i}}=\hat{\theta_{i-1}} + K_{i}(\tau_{i} - \hat{\Phi^{'}_{i}}\hat{\theta_{i-1}})
\end{equation}
\begin{equation}
K_{i}=P_{i}\hat{\Phi_{i}}=P_{i-1}\hat{\Phi_{i}}(I+\hat{\Phi^{'}_{i}}P_{i-1}\hat{\Phi_{i}})
\end{equation}
\begin{equation}
P_{i}=P_{i-1}-P_{i-1}\Phi_{i}(I+\Phi^{'}_{i}P_{i-1}\Phi_{i})^{-1}\Phi^{'}_{i}P_{i-1}
\end{equation}
The RLS parameter estimation was implemented as a part of a model controller. The figure \ref{fig:simcontrol} is actual simulink model of controller and simulated manipulator. The central block called "Interpreted MATLAB Fcn" is matlab implementation of robot inverse dynamics together with RLS parameter estimation. This block is implemented using persistent variables used to memorize one step ahead $K$, $P$ and $\theta$ values of RLS estimator.

\section{Simulation and Evaluation}

The described control strategy was tested on a MATLAB simulink model. The reference model of manipulator was powered by Peter Corke's robotics library, which is based on recursive Newton-Euler articulated body algorithm. The torque controller was computed symbolically from Lagrangian of a manipulator that had slightly different inertial parameters than reference model. The built simulink model has the following features:

\begin{compactitem}
\item $K_p$ and $K_v$ are fixed at some values, that were learned empirically by looking at optimal controllability and stability of reference model.
\item The output of controller is discretized using ZOH, at a rate of 100Hz. 
\item The reference trajectory is generated by interpolation between two joints space states
\end{compactitem}

\begin{figure}[H]
\includegraphics[width=6.5in]{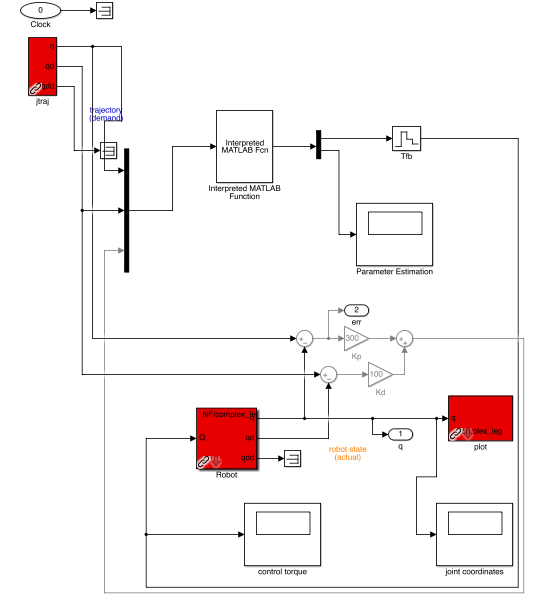}
\caption{Matlab Simulation of Controller}
\label{fig:simcontrol}
\end{figure}

The implemented simulink model is show on the picture \ref{fig:simcontrol}.

\end{multicols}
\begin{multicols}{2}


\section{Results}

The number of simulations were performed that shown that presented control strategy is viable and stable enough. In number of experiments, model parameters converged to some fixed valued. However the presented strategy has to be researched much further than the author was able to do for the project to conclude fully about effectiveness of the proposed strategy. 

\end{multicols}

\begin{figure}[H]
\includegraphics[width=6.0in]{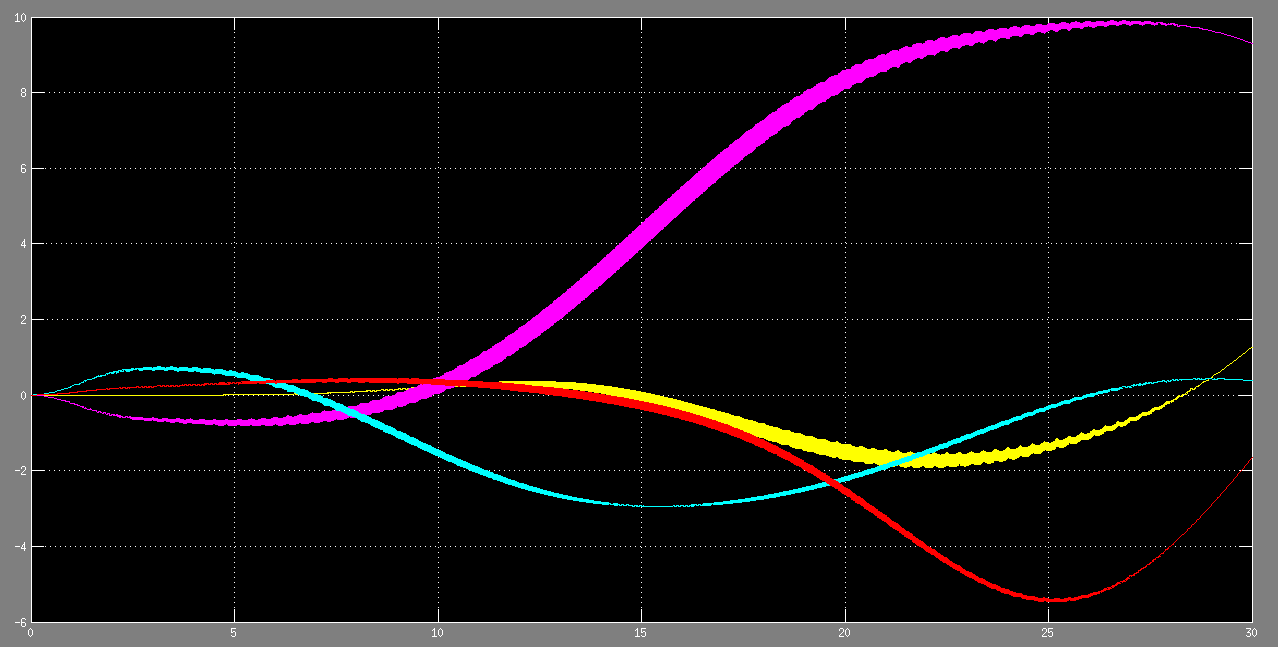}
\caption{Example output of computed torque controller}
\label{fig:simtorque}
\end{figure}

\begin{figure}[H]
\includegraphics[width=6.0in]{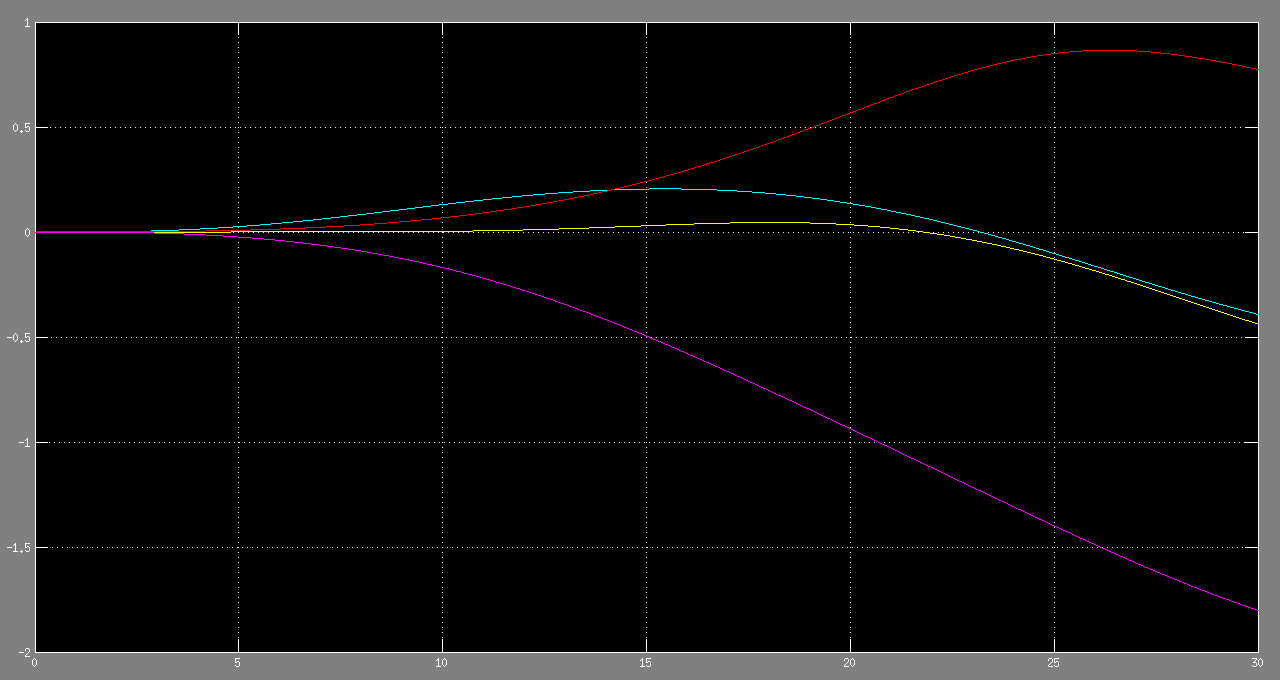}
\caption{Joint trajectories of controlled robot}
\label{fig:joints}
\end{figure}

\begin{figure}[H]
\includegraphics[width=6.0in]{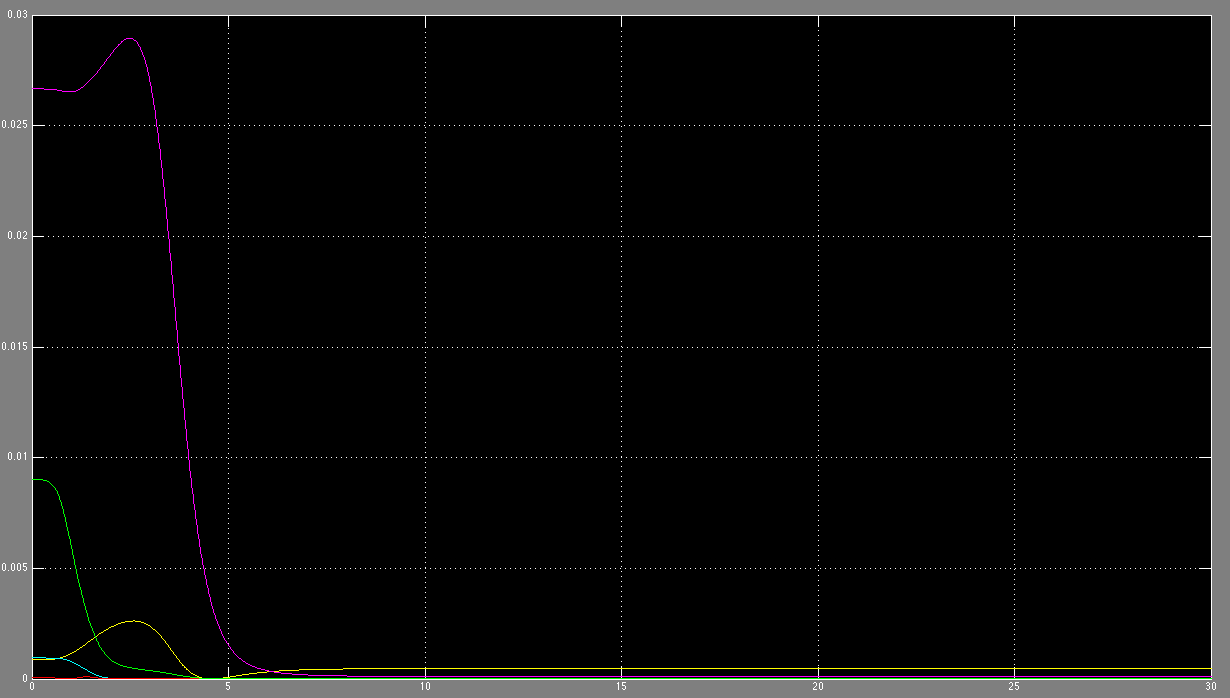}
\caption{Square error in online parameter estimation}
\label{fig:estimation}
\end{figure}

\begin{multicols}{2}



\end{multicols}

\end{document}